\documentclass[prd, preprint, nofootinbib,superscriptaddress, showpacs,11pt]{revtex4}

\usepackage{longtable}
\usepackage{amstext}
\usepackage{amssymb}
\usepackage{graphicx}
\usepackage{epsfig}
\usepackage{esint}
\usepackage{hyperref}
\hypersetup{
    bookmarks=true,         % show bookmarks bar?
    unicode=false,          % non-Latin characters in Acrobats bookmarks
    pdftoolbar=true,        % show Acrobats toolbar?
    pdfmenubar=true,        % show Acrobats menu?
    pdffitwindow=false,     % window fit to page when opened
    pdfstartview={FitH},    % fits the width of the page to the window
    pdftitle={My title},    % title
    pdfauthor={Author},     % author
    pdfsubject={Subject},   % subject of the document
    pdfcreator={Creator},   % creator of the document
    pdfproducer={Producer}, % producer of the document
    pdfkeywords={keyword1} {key2} {key3}, % list of keywords
    pdfnewwindow=true,      % links in new window
    colorlinks=true,       % false: boxed links; true: colored links
    linkcolor=red,          % color of internal links
    citecolor=cyan,        % color of links to bibliography
    filecolor=magenta,      % color of file links
    urlcolor=green,           % color of external links
    linktocpage=true
}

\makeatletter

\DeclareFontEncoding{LGR}{}{}
\DeclareTextSymbol{\~}{LGR}{126}
\providecommand{\tabularnewline}{\\}

\makeatother

\def\lt{\left}
\def\rt{\right}

\def\fr{\frac}

\begin{document}

\title{Time Delay between Relativistic Images as a probe of Cosmic Censorship}
\author{Satyabrata Sahu\footnote{Electronic address: satyabrata@tifr.res.in}}
\affiliation{Tata Institute of Fundamental Research\\
Homi Bhabha Road, Mumbai 400005, India}

\author{
Mandar Patil\footnote{Electronic address: mandar@iucaa.ernet.in}}
\affiliation{Tata Institute of Fundamental Research\\
Homi Bhabha Road, Mumbai 400005, India}
\affiliation{Inter-University Centre for Astronomy and Astrophysics\\ Ganeshkhind, Pune 411 007, India}
\author{
D. Narasimha\footnote{Electronic address: dna@tifr.res.in}}
\affiliation{Tata Institute of Fundamental Research\\
Homi Bhabha Road, Mumbai 400005, India}
\author{
Pankaj S. Joshi\footnote{Electronic address: psj@tifr.res.in}}
\affiliation{Tata Institute of Fundamental Research\\
Homi Bhabha Road, Mumbai 400005, India}

\begin{abstract}

We study the time delay between successive relativistic images in
gravitational lensing as a possible discriminator between various collapse end
states and hence as a probe of cosmic censorship. Specifically we
consider both black hole and naked singularity spacetimes admitting
photon spheres where an infinite number of relativistic Einstein Rings can be formed
at almost the same radius and naked singularity spacetimes without a photon sphere
where multiple relativistic Einstein Rings can form almost up to the
center of the lens. The metrics we have considered to sample these
scenarios are the Schwarzschild black hole, Janis-Newman-Winicour
(naked singularity with photon sphere), Joshi-Malafarina-Narayan, and Tolman-VI
(naked singularities without photon sphere) which are a fair sample
of the theoretical possibilities. We show that the differential time delay 
between the relativistic images for naked singularities without a photon sphere progressively decreases for the models we study as opposed to that for black holes and naked singularities with a photon sphere, where it is known to be roughly constant. This characteristic difference in the time delay between successive images for these two cases can be potentially  exploited for a source with known intrinsic variability to discriminate  between these scenarios even when the images are not spatially resolved.

\end{abstract}
\maketitle

\section{Introduction\label{intro} }

Gravitational lensing is a powerful astrophysical as well as cosmological
probe to study the lens object, and the source as well as the background geometry
and it indirectly serves as a probe of gravitational physics. Gravitational
lensing can help us to detect or infer the existence of exotic objects
in the universe. When the lens is a very compact object (e.g. a black
hole or the alternative end states of collapse), light rays can probe
the strong gravitational field produced by them and gravitational
lensing becomes a powerful probe of physics in strong gravity regime. If
the lens is compact enough and light rays can explore spacetime
close enough to it, then the deflection angle can become more than $2\pi$ \cite{dar59,dar61}.
Moreover propagation of light in the vicinity of black holes or compact enough objects leads to interesting signatures in the apparent position and flux of images for sources near (and orbiting) such objects \cite{Campbell:1973ys,cb72,cb73}. These signatures constitute important strong field tests of General Relativity. For faraway sources lensed by a compact object, when the deflection angle exceeds $2\pi$,  two sequences of images are formed
on different sides of the lens due to photons which undergo one or
more turns around the lens (the deflection angles are close to $2\pi,4\pi$
etc.) in addition to two (far-field) images of the source formed
due to photons which undergo a small deflection (the so-called
primary and secondary images). These additional images are  called relativistic images, the phenomenon is called relativistic lensing \cite{Virbhadra:2000ju,Virbhadra:2002ju}.

One of the exciting questions that the phenomenon of relativistic
lensing can help us address is Cosmic Censorship conjecture \cite{pen69} and the
related questions of the end state of gravitational collapse and the final
fate of massive stars \cite{IAU:8682057}. Cosmic censorship conjecture, viz. the proposal that naked singularities do not occur
in nature, still remains an open issue. 
On the other hand it has been shown that both black holes and naked singularities can form
in gravitational collapse of a matter cloud (obeying certain energy 
conditions for physical reasonability) starting from a regular
initial data \cite{Joshi1,Joshi2,Harada:1999jf}. For example, while homogeneous dust collapse always results in a black hole, inhomogeneous dust collapse with a decreasing density profile away from the center can result in both black holes and naked singularities as endstates \cite{Joshi1}. A similar conclusion follows from the study of more complex scenarios with the  inclusion of pressure. There remains still the question of whether one needs finely tuned initial data to form naked singularities. Strictly restricting to a spherically symmetric collapse, one does have a
concrete realization of naked singularity formation without the need of fine tuning initial data.
The general relativistic Larson-Penston solution, which is obtained without fine-tuning in the spherically symmetric collapse of  a perfect fluid with a soft equation of state ($p = k \rho$ for $0< k \leq 0.03$) \cite{Harada:2001nh} is stable against spherical linear 
 perturbations and describes the formation of a naked singularity  for $0 < k <0.0105$, ie for an  
 extremely soft equation of state \cite{Ori:1987hg,Ori:1989ps}. Furthermore it was shown in \cite{Joshi:2011qq} that for  spherically symmetric gravitational collapse of a general Type I matter field, when the effects of small pressure perturbations in an otherwise pressure-free collapse scenario is taken into account, both black holes and naked singularities are generic (for suitable definition of genericity) outcomes of a complete collapse. 
 %Going beyond spherical symmetry, Shapiro and Teukolsky \cite{sat91} have provided an example  where a naked singularity may form in non-spherical relativistic collapse. 
Non-spherical collapse has also been studied (see \cite{Joshi2} and references therein); but it is fair to say that a lot needs to be done before a conclusion can be arrived at for fine tuning issues in non-spherical collapse scenarios. However one can adopt the point of view alluded to in \cite{Joshi2} that `if cosmic censorship is to hold as a basic principle of nature, it better holds in spherical class too' and consider it worthwhile to investigate astrophysical implications of naked singularities. We must, all the same, mention that the examples studied here are toy models and we  hope that it captures qualitatively the basic physics so as to guide future explorations.
 
A lot of black hole candidates (compact,dark, heavy objects) have
been discovered observationally and most likely they are indeed black
holes \cite{Visser:2009pw}. However in the absence of concrete theoretical
reasons to rule out a naked singularity and in the face of our relative ignorance of the behavior of matter in an extremely high density configuration which is reached towards the end of the gravitational collapse of a massive stars, one can take a phenomenological approach,
i.e., compute the signatures of the naked singularities and confront
them with observations. Gravitational lensing signatures of naked
singularities will then be very useful in this regard. With this philosophy, Virbhadra,Narasimha and Chitre \cite{Virbhadra:1998dy} and later Virbhadra and Ellis \cite{Virbhadra:2002ju}  and Virbhadra and Keeton \cite{Virbhadra:2007kw} have studied and compared gravitational lensing by Schwarzschild black holes and by  the Janis,
Newman, Winicour naked singularities (JNW solution). Gravitational lensing
by a rotating version of the JNW naked singularity has been studied in \cite{Gyulchev:2008ff}. Extending the same line of work, we have recently studied a class of naked singularity
solutions obtained as the end state of certain dynamical collapse
scenarios for a fluid with only tangential pressure \cite{spnj12}. In this work, we take this program further to study a generalization of this model ,i.e., an analogous scenario for a fluid with non-zero radial as well as tangential pressure, to examine if the
earlier conclusions obtained for the tangential pressure case also generalize qualitatively.
Furthermore we focus on the role of time delay in relativistic lensing from this perspective by studying the time delay between relativistic Einstein Rings for  all the above mentioned solutions.
It is worth noting here that apart from lensing, accretion disk properties have also been  explored recently as a probe of cosmic censorship \cite{Kovacs:2010xm,jmn13}.

As we have discussed earlier in the strong gravity regime, 
like a black hole, the bending angle could be more
than 2$\pi$ and multiple relativistic images might be formed. In
the case of the formation of multiple images, the light-travel time along
light paths corresponding to different images is different. Therefore,
if the background source in the lens system were to be variable, this
variability will not appear simultaneously in the images. So, intrinsic
luminosity variations of the source manifests in the images as a relative
temporal phase which is expected to depend on the lens geometry and
lensing configuration. The time lag for the appearance of the intrinsic
variability between the multiple images is called the time delay.
Time delay has a privileged status among lensing observable in the
sense that it is the only dimensional observable and as such is the only
observable sensitive to the overall scaling of all the distance scales
in the problem. The observed configuration in the sky imposes dimensionless
constraints only, but time delay introduces
a physical scale in the system. That makes it a useful probe of the mass
of the lens as well as the distances between lens and observer etc. In
cosmological contexts, it has been argued to be a probe of
the Hubble parameter \cite{Refsdal:1964nw,Blandford:1991xc}. In the
conventional post-Newtonian approximation, the difference in the
light-travel time can be decomposed into a geometrical component due
to difference in the path length and a potential component due to
the different Newtonian gravitational potential felt by the photon. Time
delay between far field images is a probe of the so-called `effective
distance' of the system and can be a probe of cosmological parameters \cite{2002BASI...30..723N}. Time delay between relativistic images for black hole has been suggested as a distance
estimator \cite{bom04}. In this paper we focus on time delay as a
discriminator between black holes and naked singularities. For this we
study Schwarzschild black hole, JNW naked singularity, JMN solution and Tolman-VI solution. The last two solutions have  been shown to be obtainable as collapse end states \cite{jmn,jmn13}.  Time delay between relativistic images have previously been studied for Schwarzschild black hole in \cite{Virbhadra:2000ju} and in general for spacetimes with photons spheres in \cite{bom04} (which includes black holes and Weakly Naked Singularities, as we will see later). The present study complements these by computing time delay between relativistic images for Strongly Naked Singularities, ie., naked singularities not covered by photon sphere. To give a feel for the numbers, we consider both the case of super massive objects at galactic centers (most likely
super massive black holes \cite{2010RvMP...82.3121G}) as well as
$100M_{\odot}$ objects in our galaxy.

This paper is organized as follows. In section \ref{lf} we briefly
discuss lensing formalism and in section \ref{metric} we introduce
the spacetimes we are going to study. In section \ref{ps} we discuss
the photon spheres in these spacetimes and in section \ref{rd} we
discuss possibility of relativistic lensing and the basic features
of relativistic images. Then in section \ref{er} we discuss Einstein
Rings and section \ref{tdel} we present a discussion of time delay
between the Einstein Rings for various scenarios under consideration
and highlight the main differences between black holes and strongly
naked singularities. Finally we conclude with a discussion of main
results in \ref{dis}. We work in units of $ c=1 $ and $ G=1 $ through out the paper.

\section{Lensing Formalism\label{lf}}

In this section we briefly review the gravitational lensing formalism \cite{Virbhadra:1998dy, Virbhadra:2002ju}
We consider a spherically symmetric, static and asymptotically flat spacetime to be thought of as gravitational lens. Source and observer are assumed to be sufficiently away from the central region.
There are two important parts to the lensing formalism. First one is the lens equation which relates image
location to the source location given the deflection angle and this is essentially a geometrical relation written down taking advantage of asymptotic flatness. For a very nice discussion on various lens equations in literature, refer to
\cite{Bozza:2008ev}. The second part is the calculation of deflection angle
which is computed by integrating the null geodesics. It is through this deflection angle that General Relativity enters into picture for calculation of lensing observables such as image position and magnification.

\begin{figure}
\includegraphics[scale=0.8]{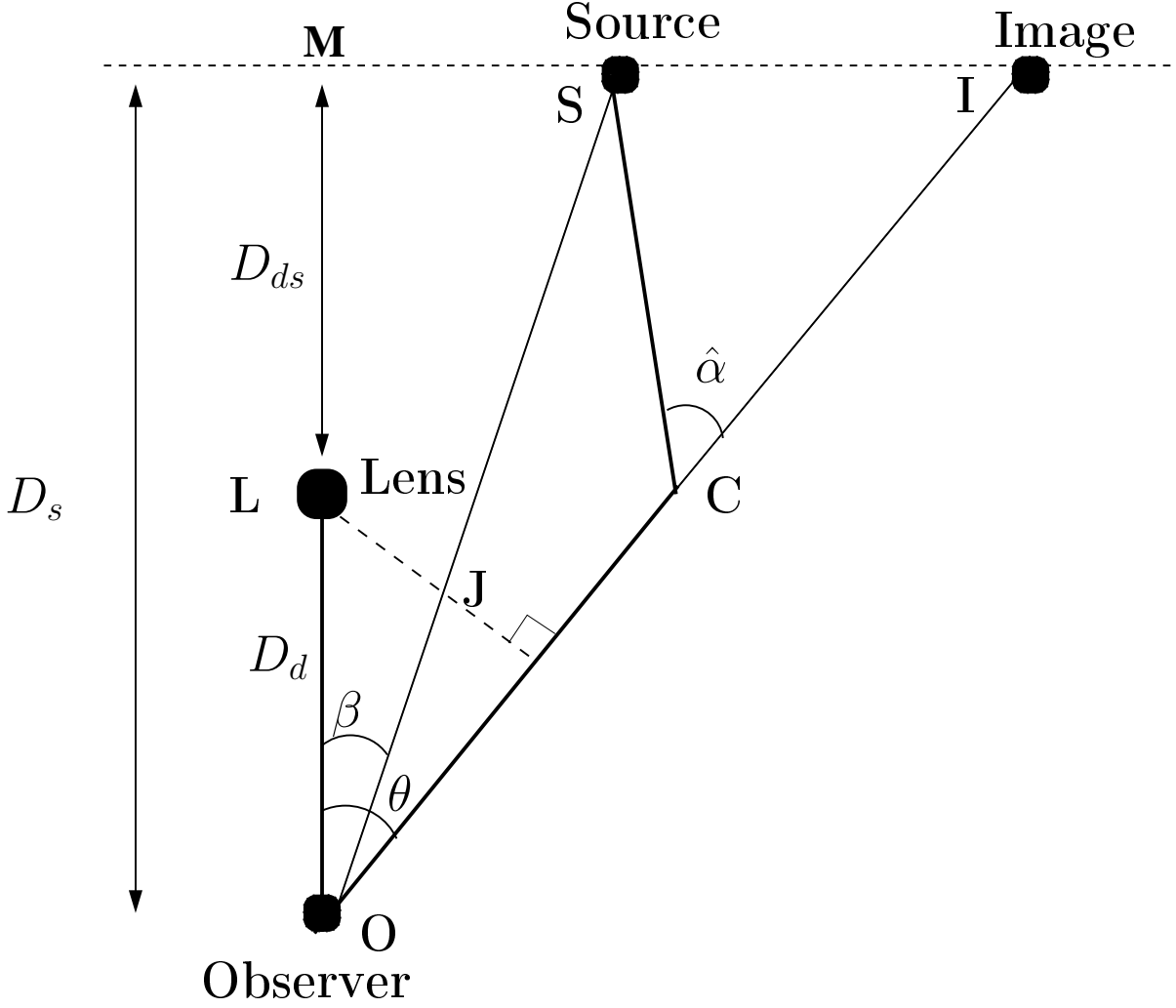}

\caption{\textit{Len Diagram}: Positions of the source, lens, observer and the 
image are given by $S$, $L$, $O$ and $I$. The distances between lens-source, 
lens-observer and source-observer are given by $D_{ds}$, $D_{d}$ and $D_{s}$. 
The angular location of the source and the image with respect to optic axis are 
given by $\beta$ and $\theta$. The impact parameter is given by $J$.\label{lens}}
\end{figure}

We use the Virbhadra-Ellis lens equation \cite{Virbhadra:2002ju} which is given by the following expression.
\begin{equation}
 \tan{\beta}=\tan{\theta}-\alpha
\end{equation}
where
\begin{equation}
 \alpha=\frac{D_{ds}}{D_s}\left[\tan{\theta}+\tan{\left(\hat{\alpha}-\theta\right)}\right]
\end{equation}
where $\theta,\beta$ denote the image location and source location respectively (see Fig \ref{lens}).
We also have $\sin{\theta}=\frac{J}{D_d}$ where $J$ is the impact parameter.

The general form of the metric is given by 
\begin{equation}
 ds^2=-g(r)dt^2+\frac{1}{f(r)}dr^2+h(r)r^2d\Omega^2
 \label{met}
\end{equation}
 The total deflection suffered 
by the light ray as it travels from source to observer is given by 
\begin{equation}
\hat{\alpha}\lt(r_0\rt)
 = 2 {\int_{r_0}}^{\infty}
\lt(\frac{1}{f(r)h(r)}\rt)^{1/2}
\lt[
\lt(\frac{r}{r_0}\rt)^2
\frac{h(r)}{h(r_0)} \frac{g(r_0)}{g(r)} -1
\rt]^{-1/2}  \frac{dr}{r}  - \pi ,
\label{defl}
\end{equation}
where $r_0$ is the distance of closest approach. The relation between impact parameter and the distance of closest approach is given by 
$J=r_0\sqrt{\frac{h(r_0)}{g(r_0)}}$.

Image location $\theta$ is obtained by solving the lens equation for the fixed value of source location 
$\beta$. The magnification is defined as 
\begin{equation}
\mu \equiv \lt( \frac{\sin{\beta}}{\sin{\theta}} \ \frac{d\beta}{d\theta} \rt)^{-1}.
\end{equation}
which can be broken down into the tangential and radial magnification in the following way.
\begin{equation}
\mu_t \equiv \lt(\frac{\sin{\beta}}{\sin{\theta}}\rt)^{-1}, ~ ~ ~
\mu_r \equiv \lt(\frac{d\beta}{d\theta}\rt)^{-1}
\end{equation}
Singularities of the tangential and radial magnification give tangential and radial critical curves and caustics. We discuss time delay separately in section \ref{tdel}.

\section{Spacetimes under consideration\label{metric}}

We will be considering static spherically symmetric spacetimes for
simplicity. Under these assumptions the simplest black hole solution is the
Schwarzschild solution which is given by

\begin{equation}
ds^{2}=-\left(1-\frac{2M}{r}\right)dt^{2}+\left(1-\frac{2M}{r}\right)^{-1}dr^{2}+r^{2}d\Omega^{2}
\label{mets}
\end{equation}
 where $M$ is the Schwarzschild mass. A naked singularity solutions which has
been studied in literature is the JNW solution \cite{Virbhadra:1998dy,Virbhadra:2002ju}.
It is a solution of Einstein equations with a minimally coupled massless
canonical scalar field and is parametrized by two parameters, mass
$M$ and scalar charge $q$. The solution can be written as

\begin{equation}
ds^{2}=-\left(1-\frac{b}{r}\right)^{\nu}dt^{2}+\left(1-\frac{b}{r}\right)^{-\nu}dr^{2}+\left(1-\frac{b}{r}\right)^{1-\nu}r^{2}d\Omega^{2}
\end{equation}
where $b=2\sqrt{M^{2}+q^{2}}$ and $\nu=\frac{2M}{b}$.

Because of the absence of analogues of black hole uniqueness theorems for naked singularities, it becomes necessary to study other possible naked singularity solutions in order to properly gauge the prospects of any observational probe (like gravitational lensing) for shedding light on cosmic censorship question. Below we describe two naked singularity solutions which, unlike the JNW solution described above, have been shown to be obtainable as collapse end states \cite{jmn,jmn13} . Admittedly these too are toy models, but  study of these solutions can be thought of as a step in the direction of study of more realistic models where by realistic we mean a naked singularity solution generated by a well motivated source, possibly taking into account deviations from spherical symmetry in the form of higher multipoles (which becomes important because of absence of analogues of black hole uniqueness theorems for naked singularities), and shown to be stable under linear (and non-linear) perturbations etc. Unfortunately such analytic solutions are hard to find or construct. One can then investigate various toy model scenarios to gather some idea of what is to be expected in a `realistic' scenario. This is expected to provide valuable insights as far as qualitative features are concerned.  JNW solution mentioned above, which is well studied in literature from lensing perspective \cite{Virbhadra:1998dy,Virbhadra:2002ju} is one such toy model as are the two solutions we describe below. One can hope that these examples capture most of the essential features of a realistic model at least for spherically symmetric situation and  observational features  would generalize qualitatively.

One of the examples is the JMN naked singularity
which was shown to be obtainable as as the end state of dynamical
collapse from regular initial conditions for a fluid with zero radial
pressure but non-vanishing tangential pressure \cite{jmn} and  was studied in \cite{spnj12} from lensing perspective. The interior region in this spacetime is described by the metric 
\begin{equation}
ds_{e}^{2}=-(1-M_{0})\left(\frac{r}{R_{b}}\right)^{\frac{M_{0}}{1-M_{0}}}dt^{2}+\frac{dr^{2}}{1-M_{0}}+r^{2}d\Omega^{2}.
\end{equation}
The solution has a naked singularity at the center and matches to
a Schwarzschild spacetime across the boundary $r=R_{b}$ with the
Schwarzschild mass given by $M=\frac{M_{0}R_{b}}{2}$. There are two parameters
in the solution, $M_{0}$ which is a dimensionless parameter and,
$M$ ehich is the Schwarzschild mass. We must have $0<M_{0}<1$ for satisfying the
condition that sound speed inside the cloud is always positive and
less than speed of light \cite{jmn}. As mentioned before, the radial pressure of the interior fluid is zero and the tangential pressure $p_t$ is related to energy desity $\rho$ as 
\begin{equation}
p_t=\frac{M_0}{4(1-M_0)}\rho
\end{equation}
Thus tangential pressure is linearly related to energy density. Both energy density and tangential pressure fall off as $1/r^2$.
In addition to these, the fourth example that we consider in this paper is  a static perfect fluid sphere solution given
by Tolman \cite{Tolman:1939jz} which also happens to be obtainable
as collapse end state \cite{jmn13}. We note here that basic
features of accretion disks in JMN and above Tolman solution was studied
 and contrasted with black hole case in \cite{jmn} and \cite{jmn13}
respectively. Below we describe the Tolman solution (henceforth to be referred to as Tolman-VI solution) briefly.

The metric in the interior is given by 
\begin{eqnarray}
ds_{e}^{2} & = & -(Ar^{1-\lambda}-Br^{1+\lambda})^{2}dt^{2}+(2-\lambda^{2})dr^{2}+r^{2}d\Omega^{2}
\end{eqnarray}
which has a central singularity and is matched to a Schwarzschild
spacetime via $C^2$ matching. The number of parameters describing the solution as written
above is 5 viz,$A,B,\lambda$ (which characterize the fluid via its
pressure and energy density) $M$ the Schwarzschild mass and $R_{b}$, the boundary
of the cloud which is the matching radius. However three matching
conditions,viz, matching of $g_{rr}$, $g_{tt}$ and pressure across
the boundary, reduce the number of independent variables to 2 and
we take $\lambda$ and $M$ as the independent variables. Then we have
\begin{equation}
R_{b}=\frac{2M(2-\lambda^{2})}{1-\lambda^{2}}
\end{equation}
\begin{equation}
A=\frac{(1+\lambda)^{2}}{4\lambda\sqrt{2-\lambda^{2}}}\left(\frac{2M(2-\lambda^{2})}{1-\lambda^{2}}\right)^{\lambda-1}
\end{equation}

\begin{equation}
B=\frac{(1+\lambda)^{2}}{4\lambda\sqrt{2-\lambda^{2}}}\left(\frac{1-\lambda}{1+\lambda}\right)^{2}\left(\frac{2M(2-\lambda^{2})}{1-\lambda^{2}}\right)^{-\lambda-1}
\end{equation}
By imposing the condition that sound speed inside the cloud is always
positive and less than speed of light ie $c_{s}<1$ one gets a bound
on physically admissible range of $\lambda$,viz, $\lambda\in(0,1)$\cite{jmn13}.
The interior fluid has pressure $p$ and energy density $\rho$ related as 
\begin{equation}
p=\frac{1}{1-\lambda^2}\frac{(1-\lambda)^2-(1+\lambda)^2r^{2\lambda}(B/A )}{1-r^{2\lambda}(B/A )}\rho
\end{equation}

which upon using matching conditions becomes

\begin{equation}
 p=\frac{(1-\lambda)^{2}}{1-\lambda^{2}}\frac{1-\left(\frac{r}{R_{b}}\right)^{2\lambda}}{1-\left(\frac{r}{R_{b}}\right)^{2\lambda}\left(\frac{1-\lambda}{1+\lambda}\right)^{2}}\rho
\end{equation} Both energy density and pressure fall off as $1/r^2$ near center and far off.

For convenience we introduce dimensionless variables (scaling all
quantities by $2M$) $x\equiv\frac{r}{2M}$, $a\equiv\frac{A}{(2M)^{-1+\lambda}},$
$b\equiv\frac{B}{(2M)^{-1-\lambda}}$. The dimensionless boundary
radius is then given by 
\begin{equation}
x_{b}\equiv\frac{R_{b}}{2M}=\frac{2-\lambda^{2}}{1-\lambda^{2}}\label{tcomp}
\end{equation}
 which means that the compactness of the solution $\frac{2M}{R_{b}}$
is determined by $\lambda$ alone. Similarly $a\mbox{ and }b$ are
also determined by $\lambda$. For convenience again we define $\kappa\equiv\frac{b}{a}$.
One then gets 
\begin{eqnarray}
a=\frac{(1+\lambda)^{2}}{4\lambda\sqrt{2-\lambda^{2}}}\left(\frac{2-\lambda^{2}}{1-\lambda^{2}}\right)^{\lambda-1}\\
\kappa=\left(\frac{1-\lambda}{1+\lambda}\right)^{2}\left(\frac{1-\lambda^{2}}{2-\lambda^{2}}\right)^{2\lambda}\label{ak}
\end{eqnarray}

\section{Photon sphere\label{ps}}

One important question for relativistic lensing is the presence/absence
of photon sphere. The deflection angle diverges in the 
limit where distance of closest approach approaches the radius of photon sphere. In fact it diverges logarithmically as demonstrated by Bozza \cite{boz02}. Photon sphere is a time-like hypersurface generated by circular closed null-geodesics in the spacetime. So it can be obtained by solving
for $r=constant$ null-geodesics or in other words for the minima/maxima
for effective potential for photons which as has already been mentioned
is given by equation \ref{pseq}. Alternatively Photon sphere can also
be defined as a time-like hypersurface $r=r_{ph}$ such that the deflection
angle becomes infinity when the closest distance of approach $r_{0}$
tends to $r_{ph}$ \cite{Virbhadra:2002ju}. The equation of the photon sphere for metric given by \ref{met} can be written as 
\begin{equation}
 \frac{g(r)'}{g(r)}=\frac{h(r)'}{h(r)}+\frac{2}{r}
 \label{pseq}
\end{equation}
where $ ' $ denotes derivative with respect to $ r $. For a generalization of
the concept of photon sphere to arbitrary spacetime see \cite{Claudel:2000yi}.
It can be argued that any spherically symmetric and static spacetime
that has a horizon and is asymptotically flat for $r\rightarrow\infty$
must contain a photon sphere \cite{Hasse:2001by} and hence any black
hole will always give rise to relativistic deflection and relativistic
images. On the other hand a naked singularity may or may not have
a photon sphere surrounding it and although a photon sphere guarantees
relativistic images, one may or may not form relativistic images in the absence of photon sphere.
Depending on whether or not the naked singularity is surrounded by
a photon sphere Virbhadra and Ellis \cite{Virbhadra:2002ju} classified
the naked singularities as weakly naked singularity (WNS) and strongly naked singularity (SNS). The ones surrounded by photon sphere are called weakly naked (WNS) while those not surrounded by photon
sphere are referred to as strongly naked (SNS). The spacetimes we study in this paper include examples of
black holes, SNS and WNS giving rise to relativistic images. We wish to clarify here that this terminology should not be confused with classification  of naked singularities into `strong curvature singularity' and `weak curvature singularity' on the basis of extendibility of spacetime through the singularity \cite{Joshi1}.

We also note here that for the  definition presented in \cite{Virbhadra:2002ju}
it is, in principle, possible to have a WNS with multiple photon spheres.
For example one could have a stable photon sphere interior to the outer unstable 
photon sphere. In such cases one can have null geodesics coming from infinity, going 
inside the photon sphere and coming back to infinity as is easy to see from the 
effective potential for photons in such spacetimes. Also there could be multiple photon sphere spacetimes \cite{Karlovini:2000xd}.
%An illustrative example of 
%such a spacetime will be Reissner-Nordstrom naked singularity for 
%$1<\frac{q}{M}<\sqrt{\frac{9}{8}}$ where $q$ is the charge parameter.
%In fact there could be regular spacetimes with two photon spheres like for example  a fluid solution matched to Schwarzschild exterior at a radius $<3M$ \cite{Hasse:2001by} and multiple photon sphere spacetimes \cite{Karlovini:2000xd}.
For such cases  lensing signature is expected to be different from the single photon sphere and  we briefly comment on this point in section \ref{rd}. The WNS spacetimes studied in this paper do not come in this category and 
always have a single photon sphere and the statements we make for WNS case 
henceforth shall consider such cases only.

For the Schwarzschild geometry the equation
of the photon sphere is satisfied at $x=3/2\mbox{ where \ensuremath{x=r/2M}}$.
In any metric matched to Schwarzschild exterior (which is the case for both
JMN and Tolman-VI metric that we are studying), the presence of photon
sphere in Schwarzschild exterior depends on the dimensionless matching radius
$x_{b}$. Clearly if and only if $x_{b}<3/2$, the exterior photon sphere will
be present. For JMN metric this condition translates to $M_{0}>2/3$.
Also as was discussed in \cite{spnj12} the interior solution has no
photon sphere when $M_{0}\neq3/2$. Thus JMN metric has a photon sphere
in the Schwarzschild exterior if $M_{0}>2/3$ and no photon sphere otherwise
\footnote{ We neglect the slightly unusual case$M_{0}=2/3$ for which there is
(neutral) photon sphere at \emph{every} radius inside the cloud.}
On the other hand for Tolman-VI metric, using $\lambda\in(0,1)$ and
\ref{tcomp} one gets $R_{b}>4M$ which implies $x_{b}>2>3/2$. So
in no regime of parameter space for this geometry one can have a photon
sphere in exterior Schwarzschild spacetime. The condition for existence of photon
sphere inside the cloud reduces to $x^{2\lambda}=-\frac{a}{b}$ and
owing to positivity of $a\mbox{ and \ensuremath{b}}$ there is no
photon sphere in the interior geometry either. Thus there is no photon sphere
in this geometry for \emph{any} value of allowed parameter range. In JNW
solution for low scalar charge $q/M\le\sqrt{3}$ one has photon sphere and hence is an example of a weakly naked singularity \cite{Virbhadra:1998dy}. In this
paper we will not be concerned with the case $q/M>\sqrt{3}$ as we
want to use JNW solution to illustrate the case of a Naked Singularity
with photon sphere. JMN for $M_{0}>2/3$ is lensing signature-wise
exactly identical to Schwarzschild as discussed in ref \cite{spnj12} and won't be considered
here. Both JMN for $M_{0}<2/3$ and Tolman-VI serve as examples of strongly
naked singularities, i.e., naked singularity without photon sphere.

\begin{figure}
\includegraphics[scale=0.8]{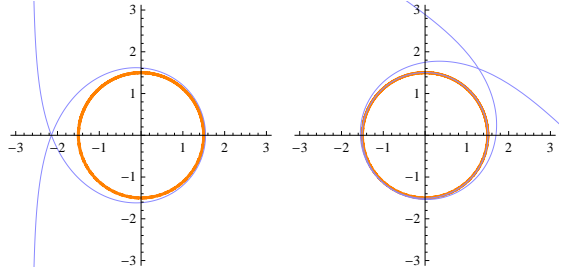}

\caption{Schematic photon trajectory for relativistic deflection in spacetime with photon
sphere with both axes plotted in Schwarzschild units (photon sphere is given by the thick orange circle; in the two-loop case the second  loop is virtually indistinguishable from the photon sphere)\label{f sch}}
\end{figure}

\begin{figure}
\includegraphics[scale=0.8]{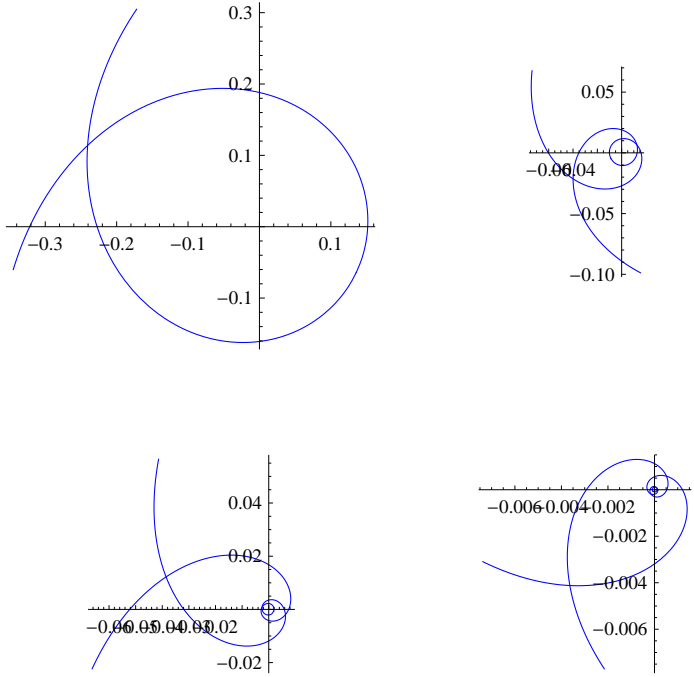}\caption{Schematic photon trajectory for relativistic deflection in spacetime without
photon sphere with both axes plotted in Schwarzschild units \label{f tol} }
\end{figure}

In figure \ref{f sch} we show schematically the photon trajectory
for photons traveling in loops in spacetime with photon sphere. As
one can see the successive loops are anchored on the photon sphere.
In contrast we show the photon trajectory for photons traveling in
loops in spacetime without photon sphere in figure \ref{f tol}. This
picture will help in an intuitive understanding of the differences in
time delay results we discuss later.

\section{Relativistic Deflection and Images\label{rd}}
Using formula \ref{defl} , the deflection angle in Schwarzschild in terms of dimensionless
variables is given as 
\begin{equation}
\hat{\alpha}\lt(x_0\rt)
 = 2 {\int_{x_0}}^{\infty}
\lt(\frac{1}{1-\frac{1}{x}}\rt)^{1/2} \lt(\frac{x}{x_0}\rt)^2
 \lt[
\lt(\frac{x}{x_0}\rt)^2
 \lt(\fr{1-\frac{1}{x_0}}{1-\frac{1}{x}}\rt) -1
\rt]^{-1/2} \ \frac{dx}{x} -\pi   \label{defls}
\end{equation}  
and 

\begin{equation}
\sin\theta=\frac{2M}{D_{d}}\frac{x_{0}}{\sqrt{1-\frac{1}{x}}}
\end{equation}
allows us to express deflection angle as a function of image location
$\hat{\alpha}(\theta)$.  For corresponding expressions in JMN refer
to \cite{spnj12}and for JNW to \cite{Virbhadra:2002ju}.
 For Tolman-VI we have for $x_{0}<x_{b}$
  \begin{eqnarray}
 \hat{\alpha}\lt(x_0\rt)=
 \nonumber
 2\int_{x_{0}}^{x_{b}}\left(2-\lambda^{2}\rt)^{1/2}\left[\lt(\frac{x}{x_{0}}\rt)^{2}
 \lt(\frac{x_{0}^{1-\lambda}-
 \kappa x_{0}^{1+\lambda}}{x_{0}^{1-\lambda}-
 \kappa x^{1+\lambda}}\rt)^{2}-1\rt]^{-1/2}
 \frac{dx}{x}\\
  +2\int_{x_{b}}^{\infty}\lt(\frac{1}{1-\frac{1}{x}}\rt)^{1/2}\lt[\lt(\frac{x}{x_{0}}\rt)^{2}\frac{a^{2}(x_{0}^{1-\lambda}-\kappa x_{0}^{1+\lambda})^{2}}{1-\frac{1}{x}}-1\rt]^{-1/2}\frac{dx}{x}-\pi\label{deflt}
 \end{eqnarray} and
\begin{equation}
\sin\theta = \frac{2M}{D_d}\frac{x_0}{a(x_{0}^{1-\lambda}-\kappa x_{0}^{1+\lambda})} \label{thx0t}
\end{equation}

where $a\mbox{ and \ensuremath{\kappa}}$ are given by \ref{ak}

It was shown in \cite{spnj12} that in JMN metric the parameter range when there
is relativistic deflection is given as $M_{0}>0.475$. Indeed for $M_{0}>2/3$
there is photon sphere and deflection angle becomes infinite. In Tolman-VI
solution however there is no photon sphere. But it turns out that the
maximal deflection becomes $>2\pi$ for $\lambda\lesssim0.44\mbox{ or equivalently \ensuremath{x_{b}<2.25}}$

The relativistic images for metric having a photon sphere are clumped
together around the angular location of photon sphere and successive
images are exponentially demagnified \cite{boz02}. This holds true
for both the Schwarzschild and the JNW metric (in an appropriate parameter range where there is a photon sphere). So there is a forbidden angular region in which no images can form and the critical angle corresponds to the angular location of the photon sphere. This is true only if there is a single (unstable) photon sphere in the spacetime. For multiple photon spheres one can have additional images corresponding to light rays that go inside the outer photon sphere and turn back and so the lensing signature is expected to be qualitatively different. While this is an interesting avenue for further study, we do not consider such cases in this paper.

In the absence of photon sphere, for example in the JMN metric
when $M_{0}<2/3$, the images are better separated and although highly
demagnified, the magnification is of the same order of magnitude for
successive images as demonstrated in \cite{spnj12} . We simply note
here that both these features qualitatively generalize for the Tolman-VI
metric despite quantitative differences and the numbers involved are
in the same ballpark. As it is not very enlightening, we forgo the
presentation of image positions and magnifications for Tolman-VI metric.
Also as with the JMN metric the deflection angle monotonically increases
with decreasing impact parameter and consequently for reasons presented
in \cite{spnj12} there are no radial critical curves in Tolman-VI solution.

\section{Einstein Ring\label{er}}

If the source, the lens, and the observer lie on a single straight
line, i.e., in the so-called aligned configuration, a circular image pattern known as
the Einstein Ring (ER), is formed. The rings formed by photons which have
been deflected by $2\pi,4\pi$ etc are referred to as relativistic
rings.
For Schwarzschild and weakly naked JNW cases, there are in principle infinite number of relativistic ERs and they are located very close to angular location of photon sphere. This is called the critical angle $\theta_{crit}$, and angular location of all images has to be greater than $\theta_{crit}$ . For a Schwarzschild black hole $4\times10^{6}M_{\odot}$ at $8.5$ kpc 
one gets $\theta_{crit}\sim24.1\text{ microarcseconds}$ and for $100M_{\odot}$ at $1$ kpc one gets $\theta_{crit}\sim5.1\text{ nanoarcseconds}$. In contrast, for SNS, ER and images are formed below the critical angle for corresponding Schwarzschild black hole, and are reasonably well separated. 
In table  \ref{ertg} and \ref{erts}, we show angular location of Relativistic ER for $4\times10^{6}M_{\odot}$ at $8.5$ kpc and for $100M_{\odot}$ at $1$ kpc  respectively for both SNS metrics studied in the paper.

\begin{longtable}[l]{|l|c|c|c|c|}
\caption{Angular location of relativistic ER for $4\times10^{6}M_{\odot}$ at $8.5$ kpc
for both SNS metrics studied in the paper:$\theta$ is in microarcseconds }
 \label{ertg}
\endfirsthead
\hline 
ER & JMN$(M_{0}=0.63)$ & JMN$(M_{0}=0.615)$ & Tolman-VI $(\lambda=0.13)$ & Tolman-VI $(\lambda=0.14)$\tabularnewline
\hline 
I & 23.96 & 23.38 & 14.90 & 14.83\tabularnewline
\hline 
II & 20.76 & 16.50 & 7.87 & 7.59\tabularnewline
\hline 
III & 15.26 & 5.96 & 4.33 & 3.89\tabularnewline
\hline 
IV & 8.17 & {*} & 1.90 & {*}\tabularnewline
\hline

\end{longtable}

\begin{longtable}[l]{|l|c|c|c|c|}
\caption{Angular location of Relativistic ER for $100M_{\odot}$ at $1$ kpc for both
SNS metrics studied in the paper:$\theta$ is in nanoarcseconds}
\label{erts}
\endfirsthead
\hline 
ER & JMN$(M_{0}=0.63)$ & JMN$(M_{0}=0.615)$ & Tolman-VI $(\lambda=0.13)$ & Tolman-VI $(\lambda=0.14)$\tabularnewline
\hline 
I & 5.09 & 4.96 & 3.16 & 3.15\tabularnewline
\hline 
II & 4.41 & 3.51 & 1.67 & 1.61\tabularnewline
\hline 
III & 3.24 & 1.26 & 0.92 & 0.82\tabularnewline
\hline 
IV & 1.73 & {*} & 0.40 & {*}\tabularnewline
\hline

\end{longtable}

In next section we compute and analyze the time delay between relativistic ERs for these spacetimes.

\section{Time Delay\label{tdel}}

The coordinate time taken by the photon to travel from $r_{0}\mbox{ to \ensuremath{r}}$
is given by 
\begin{equation}
t(r,r_{0})=\int_{r_{0}}^{r}\sqrt{\frac{1}{f(r)g(r)}}\frac{1}{\sqrt{1-\frac{h(r_{0})r_{0}^{2}}{h(r)r^{2}}\frac{g(r)}{g(r_{0})}}}dr\label{td}
\end{equation}
Let ${\cal R}_{s}$ and ${\cal R}_{o}$ be, the radial coordinates
of the source and the observer measured from the center of lens. In
dimensionless units they are given as 
\begin{equation}
\chi_{s}=\frac{{\cal R}_{s}}{2M}\text{ and }{\cal \chi}_{o}=\frac{{\cal R}_{o}}{2M}\text{.}\label{XtoR}
\end{equation}
From geometry of the configuration we can write $\chi_{s\mbox{ }}\mbox{and }{\cal \chi}_{o}$
in terms of distances and angles involved as \cite{Virbhadra:2007kw}
\begin{eqnarray}
\chi_{s} & = & \frac{D_{s}}{2M}\sqrt{\left(\frac{D_{ds}}{D_{s}}\right)^{2}+\tan^{2}\beta}\text{,}\nonumber \\
\chi_{o} & = & \frac{D_{d}}{2M}\text{,}
\end{eqnarray}
As we are concerned with ER only, we take $\beta=0$. Then as we take$\frac{D_{ds}}{D_{s}}=\frac{1}{2}$,
we have $\chi_{s}={\cal \chi}_{o}$ and equivalently $ {\cal R}_{o}={\cal R}_{s}$.
Now time difference between $m\mbox{th and \ensuremath{n\mbox{th}}}$
relativistic ER is given by 
\begin{equation}
\Delta t_{m,n}=2t({\cal R}_{0,}r_{0_{m}})-2t({\cal R}_{0,}r_{0_{n}})\label{tdd}
\end{equation}
 where $r_{0_{m}}$ is the distance of closest approach corresponding
to $m$th relativistic ER. Scaling by Schwarzschild time and writing in dimensionless
units $\tau=\frac{t}{2M}$ and for notational simplicity writing $\tau(\chi_{o},x_{0})$
as $\tau(x_{0})$ we list below we list the time delay formulas for cases under study in table \ref{tdtab}

\begin{table}
\caption{Time delay formulas (in Schwarzschild units) for spacetimes under study}

\begin{tabular}{|c|c|}
\hline 
SpaceTime & Time Delay Formulae (Dimensionless)\tabularnewline
\hline 
\hline 
Schwarzschild  & $\tau(x_{0})=2\int_{x_{0}}^{\chi_{o}}\frac{1}{1-\frac{1}{x}}\frac{1}{\sqrt{1-\frac{x_{0}^{2}}{x^{2}}\frac{\left(1-\frac{1}{x}\right)}{\left(1-\frac{1}{x_{0}}\right)}}}dx$\tabularnewline
\hline 
\hline
JMN & $\tau(x_{0})=2\int_{x_{0}}^{x_{b}}\frac{1}{\sqrt{(1-M_{0}^{2})(xM_{0})^{\gamma}}}\frac{1}{\sqrt{1-\frac{x_{0}^{2}}{x^{2}}\frac{x^{\gamma}}{x_{0^{\gamma}}}}}dx+2\int_{x_{b}}^{\chi_{o}}\frac{1}{1-\frac{1}{x}}\frac{1}{\sqrt{1-\frac{x_{0}^{2}}{x^{2}}\frac{\left(1-\frac{1}{x}\right)}{(1-M_{0})(x_{0}M)^{\gamma}}}}dx$\tabularnewline
\hline 
\hline
Tolman-VI  & $\tau(x_{0})=2\int_{x_{0}}^{x_{b}}\frac{1}{\sqrt{(1-M_{0}^{2})(xM_{0})^{\gamma}}}\frac{1}{\sqrt{1-\frac{x_{0}^{2}}{x^{2}}\frac{x^{\gamma}}{x_{0^{\gamma}}}}}dx+2\int_{x_{b}}^{\chi_{o}}\frac{1}{1-\frac{1}{x}}\frac{1}{\sqrt{1-\frac{x_{0}^{2}}{x^{2}}\frac{\left(1-\frac{1}{x}\right)}{(1-M_{0})(x_{0}M)^{\gamma}}}}dx$\tabularnewline
\hline
\hline 
JNW & $\tau(x_{0})=2\int_{x_{0}}^{\chi_{o}}\frac{1}{\left(1-\frac{1}{\nu x}\right)^{\nu}}\frac{1}{\sqrt{1-\frac{x_{0}^{2}}{x^{2}}\frac{\left(1-\frac{1}{\nu x}\right)}{\left(1-\frac{1}{\nu x_{0}}\right)}}}dx$\tabularnewline
\hline 
%\label{tdtab} 

\end{tabular}
\label{tdtab}
\end{table}

The time delay in terms Schwarzschild time is given as

$\Delta\tau_{m,n}=\tau(x_{0_{m}})-\tau(x_{0_{n}})$

Below we highlight the qualitative differences in the behavior of relativistic time delay for spactimes with a photon sphere from those without one.

\subsection{Singularities with a photon sphere (Black Holes and WNS)\label{tdps}}

As has been discussed earlier if the spacetime has a photon sphere
then there is a forbidden angular region and all relativistic images
are located close to each other and near the angular location of photon
sphere. In this case the time delay between successive relativistic
rings is more or less constant and is neatly related to light-travel
time in a circle on photon sphere. This is intuitive to understand
because in the presence of a photon sphere the light trajectories corresponding
to successive rings differ by one extra nearly circular loop at the photon
sphere radius as can be easily seen from figure \ref{f sch}. Indeed the time delay between ERs is well approximated
by 
\begin{equation}
\Delta t_{m,n}=\frac{2\pi(m-n)r_{ph}}{\sqrt{g(r_{ph})}}=2\pi(m-n)J_{ph}
\end{equation}
 where $r_{ph}$ is radius of photon sphere and $J_{ph}$ is corresponding
impact parameter \cite{Bozza:2003cp}. For Schwarzschild black hole time delay between successive
relativistic ER then becomes roughly $2\pi\times3\sqrt{3}M$ and for JNW it becomes $2\pi\times\frac{1+2\nu}{\nu}\left(1-\frac{2}{1+2\nu}\right)^{\frac{1-2\nu}{2}}M$. 
For same Schwarzschild mass, increasing scalar charge decreases the time delay. Thus
as with the case for position and magnification of relativistic images
in the presence of  photon sphere, the time delay is also determined by the metric
near photon sphere and displays certain universal features irrespective
of presence or absence of event horizon. Also as far as observations
are concerned, since in such a situation, higher order relativistic
images are clumped together (and excessively demagnified) \cite{boz02}, the relevant observational quantity should be time delay between
first and second images. The scenario becomes different in absence
of photon sphere as we discuss below.

\subsection{Singularities without photon sphere (SNS)\label{tdnps}}

\begin{longtable}[l]{|l|c|c|c|c|}

\caption{Time delay for $4\times10^{6}M_{\odot}$ at $8.5$ kpc for both SNS
metrics studied in the paper:$\Delta\tau$ is in second}
\label{tdtg}
\endfirsthead
\hline 
Time Delay & JMN$(M_{0}=0.63)$ & JMN$(M_{0}=0.615)$ & Tolman-VI $(\lambda=0.13)$ & Tolman-VI $(\lambda=0.14)$\tabularnewline
\hline 
$\Delta\tau_{2,1}$ & 299.6 & 272.0 & 145.8 & 141.9\tabularnewline
\hline 
$\Delta\tau_{3,2}$ & 242.4 & 151.8 & 78.8 & 74.9\tabularnewline
\hline 
$\Delta\tau_{4,3}$ & 157.7 & {*} & 41.4 & {*}\tabularnewline
\hline

\end{longtable}

\begin{longtable}[l]{|l|c|c|c|c|}
\caption{Time delay for $100M_{\odot}$ at $1$ kpc for both SNS metrics studied
in the paper:$\Delta\tau$ is in milisecond}
\label{tdts}
\endfirsthead
\hline 
Time Delay & JMN$(M_{0}=0.63)$ & JMN$(M_{0}=0.615)$ & Tolman-VI $(\lambda=0.13)$ & Tolman-VI $(\lambda=0.14)$\tabularnewline
\hline 
$\Delta\tau_{2,1}$ & 7.5 & 6.8 & 3.6 & 3.5\tabularnewline
\hline 
$\Delta\tau_{3,2}$ & 6.1 & 3.8 & 2.0 & 1.9\tabularnewline
\hline 
$\Delta\tau_{4,3}$ & 3.9 & {*} & 1.0 & {*}\tabularnewline
\hline 

\end{longtable}
In the absence of photon sphere the time delay between successive
images is no longer roughly constant. Explicit calculation
for various cases shows that the successive time delays go on decreasing.
In table  \ref{tdtg} and \ref{tdts}, we show time delay between Relativistic ER for $4\times10^{6}M_{\odot}$ at $8.5$ kpc and for $100M_{\odot}$ at $1$ kpc respectively for both SNS metrics studied in the paper. It remains to be seen if this is a feature generic to SNS with monotonic deflection angle. In passing we note that for the particular cases
we have studied here, the successive time
delay is lesser for the Tolman-VI case as compared to the corresponding
JMN case, where by corresponding we mean spacetimes admitting the same maximal deflection and hence the same number of relativistic images and rings. And also, the time delay is lesser in general than a black hole with the
corresponding Schwarzschild mass. In case the images are expected to be resolved
(like the super massive black hole case) the time delay between first
and last images also becomes an observational quantity of interest.
One can expect that to be fairly large if there are a large number of
images as successive time delays will add up. But the decreasing trend
in successive time delays compensates for this phenomenon to some extent.
For the super massive black hole case the time delay is of the order of seconds.
For $100 M_{\odot}$ objects at kpc distances, the time delay becomes
of the order of milliseconds. In this case images are also unlikely
to be resolved (separation as discussed earlier being of the order
of nanoarcseconds). But if there is a characteristic variability
of the source one might hope to find signatures in the unresolved image.
Relativistic Micolensing studies in such cases might be an interesting
idea to pursue.

\section{ Discussion and Conclusion \label{dis}}

In this paper, we have studied the time delay properties of gravitational
lenses in strong deflection regime when relativistic images can be formed,
and its role as an in principle probe of the cosmic censorship question.
Three of the four metrics we have considered in this paper have been
previously explored from relativistic lensing perspective. For the fourth
one viz Tolman-VI solution we find that the basic features for image positions
and magnification is similar to JMN case without photon sphere. Thus the lensing signatures for the JMN case, which happens to be a toy example of a naked singularity solution obtained as the collapse end state for a fluid with zero radial pressure, qualitatively generalize for an analogous scenario with the inclusion of radial pressure, i.e., the Tolman-VI case. We have also presented in this paper the time delay between relativistic ER for both JMN and Tolman-VI cases both of which serve as examples of SNS and contrasted them with previously studied examples for relativistic
time delay in the literature which, to the best of our knowledge, have all been either black holes or
WNS \cite{Virbhadra:2000ju,bom04,LarranagaRubio:2003vp}. We have confirmed that there are important differences between SNS and WNS for time delays between relativistic images which we discuss below. 

We have used the Schwarzschild black hole, which has a photon sphere
at Schwarzschild radius of $3M$ as a standard, against which
the time delay difference between successive Einstein Rings for our
cases of interest are compared. For the differential time delay between
relativistic images formed by successive loops made by photon, there
is practically no difference between sources along the optic axis and slightly
misaligned ones. Moreover only strongly aligned configurations are
of importance for lensing. Hence, in this work we have presented the
relativistic time delay only between the successive Einstein Rings.
This time delay is of the order of a few seconds for the galactic center
super massive object where as it is of the order of milliseconds for
100 solar mass objects in our galaxy at kpc distance scales.

For the Schwarzschild black holes, the time delay difference is very
close to $2\pi\times 3 \sqrt{3}M$,
%($2\pi\times3\sqrt{3}\frac{GM}{c^{3}}$ putting back $G$ and $c$)
which is $2\pi$ times the
critical impact parameter as to be expected. Other spacetimes with
photon sphere will emulate this feature though the time delay difference
will be different as it will depend on other parameters that characterize
the spacetime apart from mass. As an example, consider the JNW metric
discussed in section \ref{tdps} where this time delay difference decreases with
increasing scalar charge. The contrasting scenario is the case of formation of 
multiple relativistic Einstein rings in the presence of naked singularities
not covered by photon sphere. Then, the successive Einstein rings formed
for a source close to the optic axis of the lens will have successively
smaller impact parameters. The differential time delay between these
rings will progressively decrease as was shown in section \ref{tdnps}
for two prototype examples. If we were to detect time delay of this nature, possibly we have a scenario where the cosmic censorship is violated. This requires the images
to be resolved. However, even when the images are not resolved, one can possibly discriminate SNS from WNS and black holes. This is because an intrinsic source variability will manifest itself in a characteristic way in the unresolved composite Einstein Ring for black holes
and WNS owing to the (almost) periodic  nature of differential time delays (and exponential suppression of magnification for higher order rings),
which can be distinguished from the unresolved composite Einstein Ring for SNS where no such signature of periodicity is expected. Thus there are important  differences in ratio of time delays between successive relativistic images for spacetimes with photon sphere and those without photon sphere and this might be helpful in observationally distinguishing 
SNS from WNS and black holes. As we have remarked earlier, we have not considered multiple photon sphere spacetimes in this study and WNS with multiple photon spheres will have signature different from the WNS studied in this paper.

The observation of relativistic images will be a wonderful test of gravitational
physics in strong field regime. Practical difficulties and technological
challenges notwithstanding, VLBI (Very-long-baseline interferometry) is a promising technique which might in the near future achieve angular and time resolution to bring at least
some of the questions that relativistic lensing can probe such as
the one that we studied here under observational purview, at least
for the galactic central super massive object \cite{VLBI,Fish:2009va}.
With this in mind it will be useful to study more realistic examples
of naked singularities and under more realistic astrophysical conditions.

\section*{Acknowledgments}
We would like to thank Dr. K.S.Virbhadra for valuable correspondence and the anonymous Referee
for valuable comments and suggestions.

%  \bibliography{tolman}{}
%  \bibliographystyle{apsrev}

\end{document}